\documentclass[twocolumn,groupedaddress,showpacs]{revtex4}
\usepackage{amssymb}
\usepackage{graphics}
\begin{document}

\title{Giant  Relaxation  Oscillations in  a  Very Strongly Hysteretic
  SQUID ring-Tank Circuit System}

\author{T.D.~Clark}
\email{t.d.clark@sussex.ac.uk}
\author{R.J.~Prance}
\author{R.~Whiteman}
\author{H.~Prance}
\author{M.J.~Everitt}

\affiliation{Quantum Circuits Group,  School   of  Engineering,
  University of Sussex, Brighton, Sussex BN1 9QT, U.K.}

\author{A.R.~Bulsara}   

\affiliation{Space  and Naval Warfare  Systems  Centre, San Diego,  CA
  92152-5001, USA.}

\author{J.F.~Ralph}

\affiliation{Department  of Electrical  Engineering and   Electronics,
  Liverpool University, Brownlow Hill, Liverpool, L69 3GJ, UK.}


\begin{abstract}
  In this  paper  we  show  that the   radio frequency  (rf) dynamical
  characteristics of a very strongly hysteretic SQUID ring, coupled to
  an rf  tank circuit  resonator, display relaxation  oscillations. We
  demonstrate that the   the  overall form of  these  characteristics,
  together    with the   relaxation  oscillations,  can   be modelled
  accurately by  solving  the quasi-classical non-linear equations  of
  motion  for  the system. We  suggest   that in  these very  strongly
  hysteretic regimes SQUID ring-resonator systems may find application
  in novel logic and memory devices.
\end{abstract}
\maketitle

\section{Introduction}

Over  the years the   non-linear, dynamical properties of SQUID  rings
(i.e.    thick  superconducting rings  containing   either  one or two
Josephson weak  link devices)  have  formed the  basis of a   range of
technologically important devices~\cite{1},  most recently through the
phenomenon  of   stochastic  resonance~\cite{2}.   Apart   from  these
applications,  the study    of  the   physics   of   the SQUID   rings
themselves~\cite{1,3,4,5} has added considerably  to the general field
of non-linear   dynamics.    As studied previously,    these dynamical
properties  have been due to  the  quasi-classical behaviour of  SQUID
rings in the regime where quantum processes can be neglected. However,
of  late much interest  has been shown  in the  use  of SQUID rings as
purely  quantum devices  in  possible  quantum technologies~\cite{6}.  
This  interest has been   encouraged  by recent experimental  work  on
superposition   states  in weak link circuits~\cite{7,8,9,10,11,12,13}
and, in particular,  on SQUID rings~\cite{14,15}.  Inevitably, at some
stage, quantum circuits are  required to interact with classical probe
circuits  if information about their  quantum state/evolution is to be
extracted.  As we  have  shown, for the   case of the SQUID ring  this
interaction   leads to  the  growth of  non-linear   behaviour in  the
classical  part  of  the  system~\cite{5,16,17,18}.  Thus,  from  this
perspective, non-linear dynamics always has a crucial  role to play in
the description,  quantum or quasi-classical, of  SQUID rings. In this
paper  we shall demonstrate  that, even with  the quite understandable
attention  being paid to SQUID  rings  at the quantum  level, there is
much still to be explored in the quasi-classical  regime. This can add
greatly to our knowledge of non-linear systems and could lead to novel
and powerful applications.

\begin{figure}
  \resizebox*{0.48\textwidth}{!}{\includegraphics{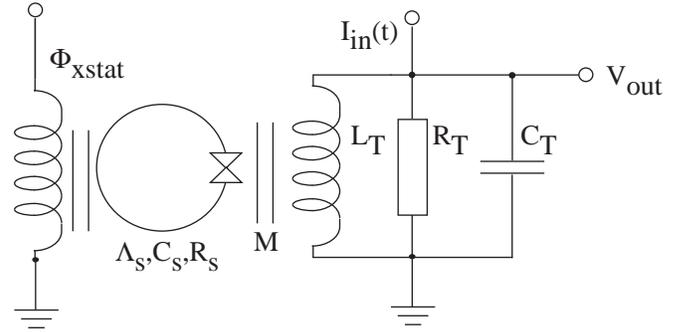}}
  \caption{
    Schematic of the  coupled SQUID ring-rf  tank circuit system  also
    showing the static magnetic flux  bias $\Phi _{xstat}$ applied  to
    the ring.  
    }
\end{figure}

In the description of SQUID rings~\cite{1} it  is customary (from this
quasi-classical viewpoint)  to identify regimes  of behaviour in terms
of the parameter $\beta \left( =2\pi \Lambda I_{c}/\Phi _{0}\right) $,
where  $ \Lambda    _{s}$ is  the   ring  inductance, $I_{c}$   is the
(Josephson) critical current  of the weak link in  the ring  and $\Phi
_{0}=h/2e$.  In the presence of an  external magnetic flux $\Phi _{x}$
the response  of a SQUID  ring  is to set  up a  screening supercurrent
$I_{s}\left( \Phi _{x}\right) $.  The  functional form of $I_{s}\left(
  \Phi _{x}\right)  $ depends on  the value of $  \beta $.   Thus, for
$\beta \leq 1$,  $I_{s}$ is always single  valued in $\Phi _{x}$ while
for $\beta   >1$ this current  is  multi-valued  and hysteretic in
$\Phi _{x}$. In this latter regime $I_{s}\left( \Phi _{x}\right) $ can
be  viewed as a    set  of approximately  diamagnetic  branches   with
switching between   adjacent   branches (in   the absence   of  noise)
occurring when $I_{\mathrm{s}}$ reaches $\pm I_{c}$. Given this branch
switching, hysteresis    loops will be   swept   out, with concomitant
dissipation, when $I_{s}$  exceeds $\pm I_{c}$.  Devices which utilize
the  properties  of    single   weak  link  SQUID   rings,   such   as
magnetometers~\cite{1}, are  usually  operated at low $\beta  $ values
(e.g.~2 to~5).  In operation as magnetometers  single weak link SQUID
rings are  typically  coupled inductively to   a  radio frequency  (rf
$\approx $ 20MHz), parallel LC, tank circuit which is excited using an
rf current  source. This    arrangement  is shown   schematically   in
figure~1. As  is well known, the rf  voltage developed across the tank
circuit is  a function of both the  rf magnetic flux amplitude  and the
static magnetic  flux $\left(  \Phi _{xstat}\right)  $ applied to  the
ring. This  static flux dependence  is periodic in  $\Phi _{0}$, hence
the use of the SQUID ring-tank circuit system in magnetometry.

The dynamical behaviour  of a   SQUID   ring coupled to an    external
resonant circuit is governed, ultimately, by the form of its potential
$U\left( \Phi ,\Phi _{\mathrm{x}}\right) $. Quasi-classically~\cite{1}
(i.e. when the effective mass of the ring -  the weak link capacitance
$C$ - is  large,  typically $10^{-12}$  to  $10^{-14}$F for  4  Kelvin
operation) this potential has the form

\begin{equation}
U\left( \Phi ,\Phi _{x}\right) =\frac{\left( \Phi -\Phi _{x}\right) ^{2}}{
2\Lambda _{s}}-\left( \frac{I_{c}\Phi _{0}}{2\pi }\right) \cos \left( 
\frac{2\pi \Phi }{\Phi _{0}}\right)   \label{squidpot}
\end{equation}
where $\Phi \left( =\Phi  _{x}+\Lambda I_{\mathrm{s}}\right) $  is the
total  included flux in  the  ring.  Clearly,  if $\Phi  _{x}$ changes
slowly with time (i.e. slowly compared  with any ring time constants),
the  ring  will always remain  at, or  very close  to,  the minimum in
$U\left(   \Phi ,\Phi _{x}\right)  $.   In dealing with ring-resonator
dynamics    it has been  customary    to  assume  that this  adiabatic
constraint  is satisfied,  e.g.    in  the  low   $\beta $  rf   SQUID
magnetometer where the peak rf flux at the ring typically $\gtrsim $ $
\Phi _{0}$ ($=\mu \varphi $ for a fraction  $\mu $ of the tank circuit
flux $ \varphi $ coupled to the ring).  The current  $I_{s}$ set up in
response to this  flux then couples back  to the tank  circuit, and so
on.  This back reaction, which  reflects $I_{s}\left( \Phi _{x}\right)
$, affects  the    rf voltage $V_{out}$   of  the  tank circuit in   a
non-linear way.   With $I_{in}$  linearly amplitude modulated  in time
(e.g.  at $\approx $ 100Hz),  the result is  the rf SQUID magnetometer
$\left(  V_{out}\  \mathrm{verses}\  I_{in}\right) $  characteristics. 
Following an initial linear riser in $ V_{out}$ versus $I_{in}$, these
characteristics consist of a set of voltage plateaux (steps) spaced at
regular  intervals along   $I_{in}$, with  period  $ \varpropto   \Phi
_{0}/\Lambda $.   The $\Phi _{0}$-periodic    modulation of these step
features by a  static (or quasi-static)  applied  flux $\Phi _{xstat}$
forms the basis of the rf SQUID magnetometer.

In     this paper we  report   on   the  experimental and  theoretical
investigation  of single weak link SQUID   rings in the so-termed very
strongly hysteretic  regime, a previously little  explored part of the
parameter space  of SQUID behaviour.   With the SQUID ring inductively
coupled to a high quality  factor $ \left(  Q\right) $ radio frequency
$\left( \mathit{rf}  \approx 20\mathrm{MHz}\right) $ resonant circuit,
we show that   quite remarkable non-linear phenomena (large  amplitude
relaxation  oscillations) can  develop.   We demonstrate  that exactly
these dynamics  are solutions of the  coupled  non-linear equations of
motion of the system.

\section{SQUID ring-tank circuit dynamics in the strongly hysteretic regime}

\subsection{Quasi-classical equations of motion}

The  quasi-classical description of the   ring-tank circuit system  is
based   on the resistively   shunted junction plus capacitance (RSJ+C)
model of a weak link. Here,  the supercurrent channel through the link
is in parallel with a normal  current channel (resistance $R_{s}$) and
a capacitance $C_{s}$.  In the model the normal  channel turns on when
the link current $I>I_{c}$. With representative values of $C_{s}$ from
$10^{-12}$ to $10^{-13}$F,  generally accepted resistances $R_{s}$ are
in the $10$ to  $100\Omega $ range~\cite{19}.  With  this in mind, the
coupled ring-tank circuit equations of motion are~\cite{1,20}
\begin{widetext}
\begin{equation}
\mathrm{SQUID\ ring}:C_{s}\frac{d^{2}\Phi }{dt^{2}}+\frac{1}{R_{s}}\frac{
d\Phi }{dt}+I_{c}\sin \left( \frac{2\pi \Phi }{\Phi _{0}}\right) +\frac{\Phi 
}{\Lambda _{s}\left( 1-K^{2}\right) }=\frac{\mu \varphi }{\Lambda _{s}\left(
1-K^{2}\right) }  \label{squidmot}
\end{equation}

\begin{equation}
\mathrm{Tank\ circuit:\ }C_{T}\frac{d^{2}\varphi }{dt^{2}}+\frac{1}{R_{T}}
\frac{d\varphi }{dt}+\frac{\varphi }{L_{T}\left( 1-K^{2}\right) }
=I_{in}\left( t\right) +\frac{\mu \Phi }{\Lambda _{s}\left( 1-K^{2}\right) }
\label{tankmot}
\end{equation}
\end{widetext}
for  tank circuit  capacitance,   inductance and   parallel  resonance
resistance  $ C_{T}$,   $L_{T}$ and   $R_{T}$, respectively,  coupling
constant $K=\sqrt{  M^{2}/L_{T}\Lambda  }$ for mutual  inductance $M$,
and Josephson current $ I=I_{c}\sin \left( 2\pi \Phi /\Phi _{0}\right)
$. It is apparent in (\ref{squidmot}) that  there are several possible
ring time constants $\left( C_{s}R_{s},\Lambda _{s}/Rs,1/\sqrt{\Lambda
    _{s}C_{s}}\right) $.  Clearly,  if the adiabatic constraint holds,
and there  are no  time  constant problems,   the derivative  terms in
(\ref{squidmot}) can be neglected and (\ref{squidmot}) reduces to

\begin{equation}
I_{\mathrm{c}}\sin \left( 2\pi \Phi /\Phi _{0}\right) +\Phi /\Lambda
_{s}\left( 1-K^{2}\right) =\mu \varphi /\Lambda _{s}\left( 1-K^{2}\right) 
\label{adcon}
\end{equation}
In   the  past this,   together  with (\ref{squidmot}),   has been the
starting point   of most dynamical  descriptions  of ring-tank circuit
systems, at least for low  $\beta $ SQUID rings.  We note that in this
linearized, small $\beta $, description $I_{c}$ (i.e. $\beta $) can be
found  from   the point of   onset  in $I_{in}$ of  the  periodic step
features  in $V_{out}$  versus   $I_{in}$ (the breakpoint)  beyond the
initial linear riser~\cite{21}.

As we emphasized recently~\cite{20},  the adiabatic constraint  breaks
down when the ring  is strongly hysteretic and  underdamped~\cite{22}. 
Qualitatively, at 4 Kelvin, transition widths for branch switchings in
$I_{s}\left( \Phi _{x}\right) $ are typically $\approx \Phi _{0}/100$.
At 20MHz, and at the rf drive levels  used experimentally for a $\beta
\approx    50-100$  ring ($\mu \varphi   \gtrsim   100\Phi _{0}$), the
available   time  to switch between  branches  $\approx 10^{-12}$secs,
comparable to the shortest ring time constant. In this range of $\beta
$  novel  multiple  level    structures develop in    the experimental
$V_{out}$  versus $I_{in}$ characteristics~\cite{20}. These structures
(plateaux of almost constant $V_{out}$  along $I_{in}$) are on a scale
in $I_{in}$ large compared with standard rf  SQUID magnetometer steps. 
However,  these  still maintain a  form  of $\Phi _{0}$-periodicity in
$\Phi _{xstat}$. We  also demonstrated numerically that these multiple
level structure characteristics  formed a new  and particular class of
solutions of ( \ref{squidmot}) and (\ref{tankmot}). Given the range of
non-linear phenomena  that  have already been recognized  in ring-tank
circuit systems,  it  would  be  surprising  if  these multiple  level
structures constituted  the only  new class of   solutions in the high
$\beta $ regime. At  even higher $\beta $  a second type of non-linear
dynamics  (also a   solution  of  \ref{squidmot}  and  (\ref{tankmot})
develops - relaxation oscillations in rf  voltage against current - on
scales  very large $\left(  \approx \times  10\right) $ compared  with
conventional SQUID magnetometer characteristics.

\subsection{Experimental $V_{\mathrm{out}}$ versus $I_{\mathrm{in}}$
characteristics}

In our experiments  we made use   of Zimmerman, 2-hole, niobium  point
contact SQUID rings~\cite{23}.  The  point contact weak links in these
rings  were        adjusted,    in    situ,    at     liquid    helium
temperatures~\cite{17,20}, a technique   which allows us  to make weak
links  in a controlled way  over a very wide range  of $\beta $ values
$\left( \mathrm{e.g.\ }1<\beta  <100\mathrm{\   }\right)   $.    In  our
experimental arrangement the tank circuit voltage $V_{out}$ (figure~1)
was first  amplified by  a liquid  helium cooled  GaAsFET preamplifier
(gain $\simeq $  20dB, noise temperature $<$  10 Kelvin).  This signal
was further boosted using a receiver of our  own design and then diode
detected.   Overall,  the   receiver system    (including the  GaAsFET
preamplifier) was designed to  combine very low  noise with very large
dynamic range and slew rate,  all three  properties being required  to
observe the very strongly hysteretic behaviour reported here.

\begin{figure}[thb]
   \resizebox*{0.48\textwidth}{!}{\includegraphics{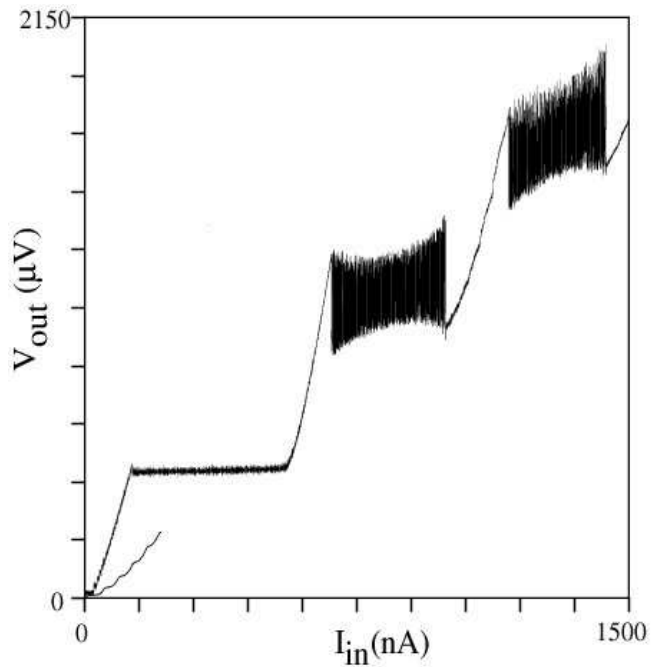}}
  \caption{
    Experimental  (4.2   Kelvin,        $\Phi  _{\mathrm{xstat}}=n\Phi
    _{\mathrm{o     }}$, $\Lambda       =6\times           10^{-10}$H,
    $L_{\mathrm{t}}=6.3\times    10^{-8}$H,        $f_{      \mathrm{\ 
        rf}}=22.671$MHz, $K^{2}=0.008$ and $Q=515$) $V_{\mathrm{out}}$
    versus  $I_{\mathrm{in}}$  characteristic   for  a   niobium point
    contact SQUID  ring-rf tank   circuit system  in the   very highly
    hysteretic  regime showing large  scale  relaxation  oscillations. 
    Here,  the triangular  ramp amplitude  modulation of  the rf is at
    $18$Hz and the detection bandwidth of the receiver is 1MHz.  
    }
\end{figure}

\begin{figure}[thb]
   \resizebox*{0.48\textwidth}{!}{\includegraphics{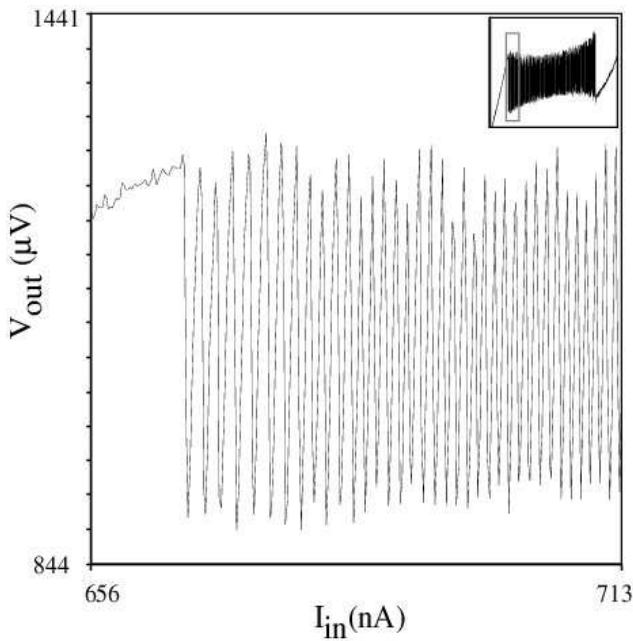}}
  \caption{
    Expanded region of the second plateau  in figure~2 showing details
    of the relaxation oscillations.  
    }
\end{figure}

In figure~2 we show a perfectly  typical  experimental diode detected
$V_{out}$ versus  $I_{in}$ SQUID  magnetometer characteristic  in this
very strongly  hysteretic regime. This was taken  at 4.2 Kelvin,  in a
bandwidth of $1$MHz, with   $Q=515$, $K^{2}=0.008$ and a  static  bias
flux  $\Phi  _{xstat}=n\Phi _{0}$,  $n$ integer,  supplied by a second
coil coupled to the ring (figure~1).  Here, adopting common practice,
this characteristic was recorded   at the $ n\Phi _{\mathrm{o}}$  bias
frequency $f_\mathit{rf}$ ($22.671$MHz in this  case)  equal to the  resonant
frequency of the ring-tank  circuit  system when $\Phi  _{xstat}=n\Phi
_{0}$.  As is  standard   for  these dynamical   (SQUID  magnetometer)
characteristics, the  rf is amplitude modulated  using a low frequency
triangular ramp. In this example the  ramp frequency is $18$Hz. As can
be seen, the characteristic consists of a first  feature, a flat step,
followed by two  further step features  in  which it  is apparent that
$V_{out}$ oscillates very rapidly as  $I_{in}$ increases. Although not
shown  in figure~2,  these oscillatory  step structures,  periodic in
$I_{in}$,  continued  to repeat as  $I_{in}$  was made  larger. The rf
voltage and current  scales in figure~2  are made clear by the  small,
subsidiary characteristic  in the same figure.  This  was taken at 4.2
Kelvin, with $\Phi _{xstat}=n\Phi _{0}$, using the SQUID ring and tank
circuit of the main characteristic but with the weak  link in the ring
adjusted to give a $\beta $ just greater than unity, i.e.  the regular
rf SQUID magnetometer  regime. It is  obvious that the scalings of the
main  characteristic are very   large indeed compared   to those for a
standard SQUID magnetometer. In figure~3 we show details of the second
step structure of  figure~2 along  which the large  scale oscillations
first appear.  Here,  we have expanded the first  section of this step
structure (shown  boxed in figure~3).  In the  expanded picture it is
apparent that $V_{out}$  is undergoing rapid oscillations  as $I_{in}$
is  increased. Furthermore, the  spacing in  $I_{in}$ between adjacent
oscillations  decreases monotonically as $I_{in}$  is swept across the
step; this  continues right across  the step (not  shown in figure~3). 
We note  that  the effect  of a  bias flux  $\Phi  _{xstat}$ on  these
oscillatory step structures was  minimal, only changing the  amplitude
of the oscillations  very weakly, with a periodicity  of  $\Phi _{0}$. 
Experimentally, a search of the first (flat) step  in figure~1 did not
reveal any voltage   oscillations, only noise.   Again,  the effect of
$\Phi _{xstat}$ was found to be negligible.

The waveform of figure~3 reflects  a limit cycle  consisting of a very
slow build up (at least on the time scale  of one rf oscillation $2\pi
/\omega _\mathit{rf}$) followed  by an extremely  fast discharge with  energy
drawn  from the tank  circuit.  This is  the followed by the next slow
build up with energy supplied from the  rf current source over many rf
periods, and so on.These are, of course, relaxation oscillations which
can be found in many  physical situations where non-linear dynamics is
involved~\cite{28,29,30,31}.  For  the  particular case of the  driven
SQUID  ring-tank circuit system of figure~3, the peak rf amplitude is
increasing  (linearly ramped with time)  from left to right. Since the
energy is made up from the rf current drive, this  means that the time
interval   between   successive  oscillations  shortens   as  $I_{in}$
increases, as is apparent in figure~3.

Without the combination of very low noise receiver (including a liquid
helium  cooled  GaAsFET  preamplifier stage)   and very  large dynamic
range/slew rate   capability, the   relaxation oscillations  shown  in
figures~2 and~3 would not be observed.  For example, reduced bandwidth
or  added noise,  or   a combination  of  both,  can  wash out   these
oscillations. What then remain are large (compared with ordinary SQUID
magnetometer characteristics - see insert  figure~2), roughly constant
rf voltage, current steps. This  is presumably why these dynamics have
not been observed in the past.

It is interesting to note that the relaxation oscillations observed by
us  in coupled SQUID  ring-tank circuit  systems  in the very strongly
hysteretic regime appear to  have   analogues in the phase    slippage
experiments  which   have    been performed  on   superfluid    helium
systems~\cite{25,26,27}. It seems clear  that, putting the fundamental
description of superfluid  and superconducting condensates aside,  the
phase/flux  slippage  in  the    two systems   is analogous.    At   a
phenomenological level this     means that the  equations    of motion
governing the dynamics   of the two   ring-resonator systems are  also
analogous.  In  the highly  hysteretic  regime for phase   slippage in
superfluid rings containing  weak  link  orifices we would   therefore
expect  to see (giant) relaxation  oscillation phenomena and, perhaps,
multilevel dynamics~\cite{20}.

\subsection{Numerical calculations}

The  $V_{\mathrm{out}}$  versus  $I_{\mathrm{in}}$ characteristics  of
figures~1    and~2 can  be    modelled  using   (\ref{squidmot}) and
(\ref{tankmot})  which can   handle  the problem  of  non-adiabaticity
perfectly   well~\cite{20}.   However, since   for  typical SQUID ring
parameters  ($C_{s}\approx     10^{-13}$F,    $\Lambda    _{s}=6\times
10^{-10}$H, $R_{s}\approx 10\Omega $), the ring time constants will be
2  or  more orders of magnitude  shorter   than the tank  circuit time
constant $\left( \cong  1/20\mathrm{MHz}\right) $, great  care must be
taken  to   ensure  that  accurate  solutions    are found.   To solve
(\ref{squidmot}) and  (\ref{tankmot}) we used fourth order Runge-Kutta
with an  adaptive step  size algorithm~\cite{5,17,20}.   We  were also
careful allow a sufficient number of steps per  rf cycle to follow the
changes  in  the  SQUID  ring  accurately.   For  our  comparison with
experiment we  set the rf  drive frequency at the $\Phi _{xstat}=n\Phi
_{0}$           resonant        frequency      (the     pseudoresonant
frequency~\cite{5,17,20}). At this bias flux  the ring is at its  most
diamagnetic [with a  ring  magnetic susceptibility $\chi  \left( n\Phi
  _{0}\right) =\partial I_{s}\left(  n\Phi _{0}\right) /\partial  \Phi
_{x}\cong -1$] and $I_{s}$ is  almost linear in  $\Phi _{x}$. From the
data of figure~2 we can make an  estimate of $I_{c}$ (and hence $\beta
$)    from  the start   of the  step  features  in  the standard SQUID
magnetometer compared with the (first) step feature in the large scale
characteristics~\cite{20}. This yields an $I_{c}$ for the large $\beta
$ characteristic $\cong 80\mu $A (i.e.  a $\beta $  value in the 70 to
80 range). We have shown~\cite{24} that knowing  $I_{c}$ and the Fermi
velocity of  the  material of the   SQUID ring (here   niobium) we can
estimate $C$.  For $ I_{c}\simeq 100\mu  $A this gives $C_{s}\approx $
few$\times   10^{-13}$F.   In   addition,   from   singly    connected
current-voltage characteristics for point contact weak links$^{19}$, a
typical value for $R_{s}$ at this $I_{c}$ is $ \approx 10\Omega $.

\begin{figure}[!thb]
   \resizebox*{0.48\textwidth}{!}{\includegraphics{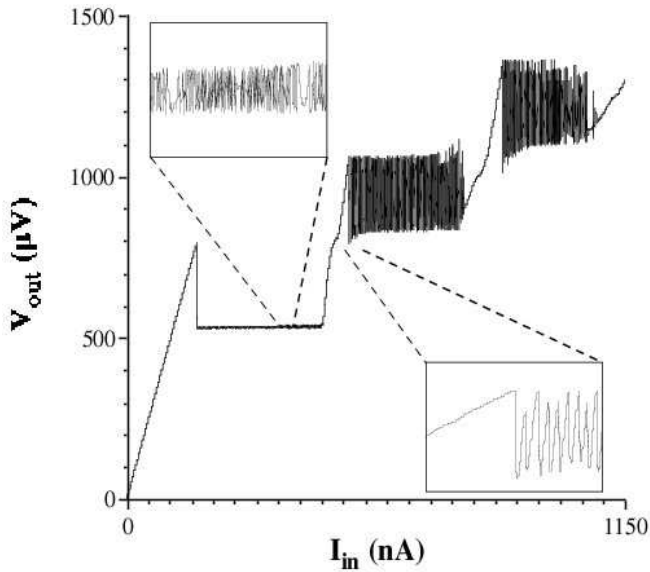}}
  \caption{
    Theoretical    (RSJ+C) $V_{\mathrm{out}}$ versus $I_{\mathrm{in}}$
    characteristic  (with a 4  Kelvin tank circuit noise source added)
    to model    the  data  of    figures~2  and~3,   setting $\Phi
    _{\mathrm{xstat}}=n\Phi     _{\mathrm{    o}}$,           $\Lambda
    _{\mathrm{s}}=6\times    10^{-10}$H,     $L_{\mathrm{t}}=6.3\times
    10^{-8}$H, $K^{2}=0.008$,  $Q=515$  and  a  best fit $f_{\mathrm{\ 
        rf}}=25$MHz.   Here,  the effective triangular  ramp amplitude
    modulation frequency is close to $30$Hz.  
    }
\end{figure}

\begin{figure}[!thb]
   \resizebox*{0.45\textwidth}{!}{\includegraphics{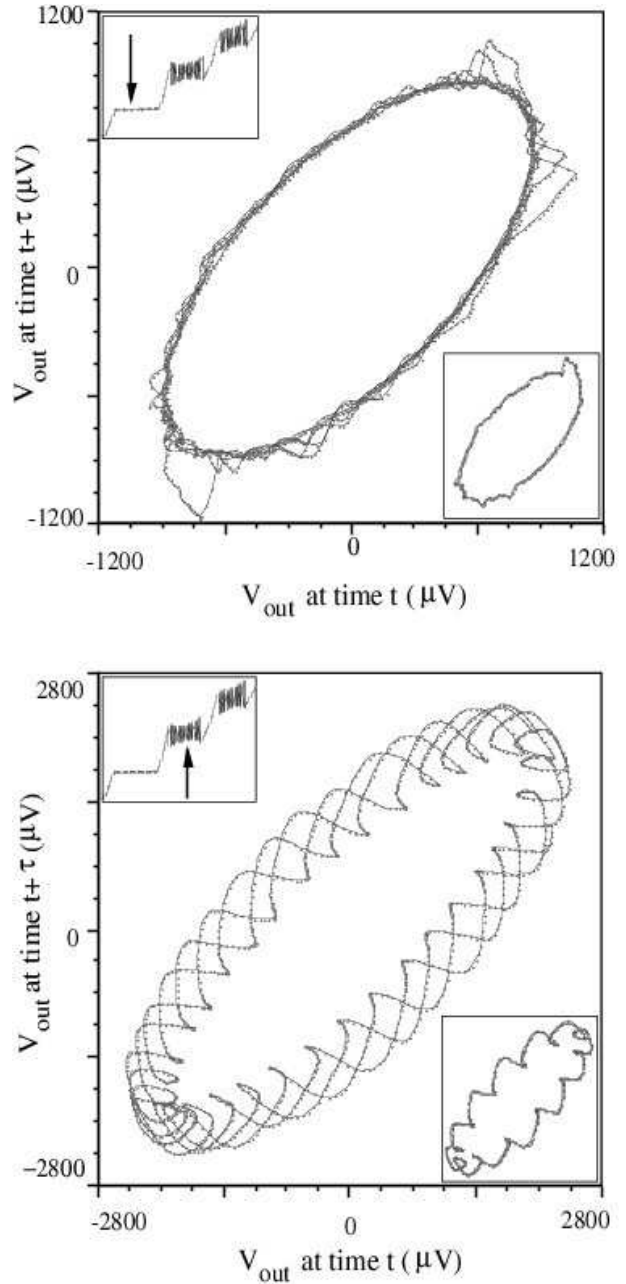}}
  \caption{
    Experimental pseudo  phase    portraits   ($V_{\mathrm{out}}\left(
      t+\tau  \right)  $, taken over sets   of complete orbits, versus
    $V\left( t\right) $) for the two $I_{ \mathrm{in}}$ bias points of
    figure~2 as shown arrowed in the  insets. Here, $\tau =10^{-4}\sec
    $ and $T=4.2$Kelvin.  The  lower insets  in  (a) and  (b)  display
    single orbits of the data sets  for the two  bias points. The data
    was  obtained in  a detection  bandwidth of  $1$MHz with an $18$Hz
    sweep frequency.  }
\end{figure}

In figure~4 we show the  results of solving (\ref{squidmot}) and (\ref
{tankmot}) to simulate the  behaviour of the ring-tank circuit  system
of figures~2 and~3, including a  4 Kelvin noise source~\cite{5,17,20}. 
Given the very  non-linear  nature of   this coupled  system,  and the
number  of parameters involved,  an   exact match  between  theory and
experiment has proved   difficult.   Nevertheless,  we were  able   to
reproduce  the important features of the   integer bias flux $V_{out}$
versus $I_{in}$ characteristic of figures~2 and~3 with $I_{c}=80\mu $A
and best fit values $C_{s}=5\times  10^{-13}$F, $ R_{s}=10\Omega $ and
$f_\mathit{rf}=25$MHz.  The  other  circuit  parameters  are,  as  for
figures~2  and~3,  $Q=515$,  $K^{2}=0.008$,  $L_{T}=63n$H, while   the
effective triangular ramp frequency (converted from an rf current ramp
in $  \mu  \mathrm{A}/\mathrm{sec}$) is  close to  $30$Hz.  As can  be
seen, the computed solution contains both the first  flat step and the
relaxation oscillations on the  second (and subsequent)  step features
of figure~2.   Details of these  (computed)  oscillations are shown in
the lower inset in figure~4.  Also shown  inset, over a small range of
$I_{\mathrm{in}}$ in the same figure, is a  highly expanded section of
the first flat step.   This  too displays (weak) voltage  oscillations
which    appear to be     either  aperiodic or   multiperiodic.  These
oscillations were  not   seen experimentally,  presumably  because  of
limited signal to  noise.    If $I_{in}$  is   fixed in   the  regions
corresponding to  the insets in figure~4 -  for  the experimental data
set  of figure~2,  and  time  series data   are collected,  these weak
oscillations can be seen reflected in  the (pseudo) phase portrait, as
shown in figure~5(a).
  Here, the pseudo-phase portrait is defined as a
plot  of $V_{out}\left(   t\right) $    against  the  time  shifted  $
V_{out}\left( t+\tau  \right) $ at all times  $t$, where $\tau $  is a
suitable  time  delay    $\left(      10^{-4}\sec \mathrm{in\    this\ 
    example}\right)  $.  In the main part  of figure~5(a)  we show the
pseudo  phase portrait, taken  over a set  of complete orbits, for the
bias point in the upper inset.  In the lower inset we display just one
of  these  orbits.  It  is   clear,  for  the   first step, that  this
pseudo-phase   portrait   simply  shows  random  fluctuations  (noise)
superimposed  on a singly periodic   orbit.  This should be contrasted
with the pseudo-phase portrait of figure~5(b) for $I_{in}$ biased onto
the  middle of the   relaxation oscillations in  the second  $V_{out}$
versus  $I_{in}$   step  feature  [shown  in    the   upper inset   of
figure~5(b)].  Here,  we  have again plotted a  set  of orbits, with a
single orbit of this set shown in the lower inset. As is apparent both
the set and the single orbit display the periodic behaviour associated
with the relaxation oscillations on the step.

What the  theoretical computations of figure~4, the experimental data
of    figures~2 and~3,  and     the previously published  multilevel
structures~\cite{20}, demonstrate is that the form of the solutions to
the coupled  equations of motion (\ref{squidmot})  and (\ref{tankmot})
is critically dependent  on parameter  values.  More specifically,  it
appears that  in the very strongly  hysteretic regime and  above (i.e. 
$\beta \gtrsim 100$) the underdamping in  the ring-tank circuit system
becomes    sufficiently large   to   wash  out   multilevel structures
completely, or almost completely.   Physically, these structures arise
(with stochastic jumping) from the system moving around $n$ hysteresis
loops in $\Phi $  versus $\Phi _{x}$ (or  $I_{s}$ versus $\Phi _{x} $)
in $m$ rf  cycles, $n$  and  $m$ integer~\cite{22}. When  sufficiently
underdamped we assume that the system simply jumps between much larger
values  of $\Phi $ in  $\Phi  $ versus  $\Phi  _{x}$, followed by long
recovery  period over  many  rf cycles, i.e.  a relaxation oscillation
process.

\section{Conclusions}

The experimental data of figures~2 and~3 are  typical of the $V_{out}$
versus $I_{in}$ characteristics found in the very strongly hysteretic,
and underdamped, regime $\left( \beta >100\right) $. These data can be
modelled    with    considerable    accuracy    invoking   the   SQUID
potential~(\ref{squidpot})  and using the quasi-classical equations of
motion~(\ref{squidmot}) and (\ref{tankmot}), as  can be seen from  the
computed $V_{out}$ versus  $I_{in}$ characteristics of figure~4. Thus,
from experiment and theory we find  that there exists another class of
solution, to the  dynamics of the driven  ring-tank circuit  system in
the    very   highly   hysteretic  regime,      apart from  multilevel
structures~\cite{20}.   Since these (latter)  structures clearly  hold
out the real  possibility  of  quasi-classical multilevel logic    and
memory if  these dynamics are well understood  and  can be controlled,
there is every  reason for opening up   the phase space  of behaviour. 
Indeed,  the potential   for single  flux  quantum  logic  has already
attracted  a great deal  of interest~\cite{32}.   It seems reasonable,
therefore, that  multilevel logic based  on the manipulation  of sets of
discrete flux states of a SQUID ring in the strongly hysteretic regime
will also  generate interest.  From the  work reported here,  it would
appear  that the  change   over   between multilevel and    relaxation
solutions is  quite subtle and  sensitive to circuit parameters.  From
experiment   and   theory   we    find  that   relaxation  oscillation
characteristics, very similar to those reported here, carry on to very
much higher $\beta  $ values $\left(  >500\right) $.  This, of course,
points to    these being a    rather  general class  of   solutions to
(\ref{squidmot})  and  (\ref{tankmot}).  Again,  as with   the earlier
multilevel structures~\cite{20}, these  solutions do not  arise in the
linearized description of the system.  We  note, again from experiment
and  computation,  that the multi-level structures~\cite{20}  found in
the strongly hysteretic regime  with $\beta \approx 50$ gradually fade
away  as   $\beta $ increases,  to   be  replaced   by the  relaxation
oscillation characteristics reported in  this paper.  In our  opinion,
even though other factors may   be important in governing the  already
underdamped dynamics of this highly non-linear  system, this points to
the damping becoming  weaker as $\beta $ grows.   It also  seems clear
that a rich new field of non-linear  dynamics exist which, as yet, has
only been partially  explored.  From a  historical perspective, it  is
also of interest  that the  original  description of rf-biased   SQUID
magnetometers~\cite{21}  was based on  a relaxation oscillation model,
albeit operating over  very  much  shorter  time intervals  (a few  rf
periods),  and on  the rf  voltage and current  scales of conventional
SQUID magnetometers~\cite{21}.  Never the less, relaxation oscillations
have  not been seen in the  dynamics of these  standard magnetometers. 
To our best knowledge,   the relaxation oscillations reported in  this
paper are the first to have been observed  in the dynamical ($V_{out}$
versus  $I_{in}$) characteristics  of SQUID  ring-resonator systems on
any scale or frequency.
 
\section{Acknowledgements}

We   would to  express our thanks   to the   Engineering and  Physical
Sciences Research Council for its generous funding of this work.

\end{document}